\RequirePackage{fixltx2e}
\documentclass[aip,jcp,reprint,floatfix]{revtex4-1}

\usepackage{amsmath,amsfonts} %
\usepackage{wasysym} %
\usepackage[acronym]{glossaries} %
\usepackage{graphicx} %
\usepackage{dcolumn} %
\usepackage{bm} %
\usepackage{ifpdf} %
\usepackage{color} %
\usepackage{nicefrac} %
\usepackage{natbib} %
\usepackage[byname]{smartref} %
\usepackage[breaklinks, %
colorlinks, linkcolor=blue, citecolor=blue, urlcolor=blue]{hyperref} %
\usepackage[table]{xcolor} %
\usepackage{braket}

\DeclareMathOperator{\tr}{Tr} %
\newcommand{\braOket}[3]{\ensuremath{\left\langle #1\left|#2\right|#3\right\rangle}}

\newcommand{\op}[1]{\ensuremath{\hat{#1}}}

\newcommand{\mat}[1]{\ensuremath{\boldsymbol{#1}}}
\newcommand{\dt}[1]{\dot{#1}}
\newcommand{\Dim}[0]{\mathcal{D}}
\newcommand{\Hilb}[0]{\mathcal{H}}
\newcommand{\inumb}[0]{\dot{\imath}}

\newcommand{\Nonunit}[0]{\mathcal{N}}
\newcommand{\sqrtNonunit}[0]{\mathcal{O}}

\ifpdf %
\fi

\begin{document}
                            
\newacronym{CI}{CI}{conical intersection} %
\newacronym{LVC}{LVC}{linear vibronic coupling} %
\newacronym{DOF}{DOF}{degrees of freedom} %
\newacronym{PES}{PES}{potential energy surface} %
\newacronym{BMA}{BMA}{bis(methylene) adamantyl} %
\newacronym{FC}{FC}{Franck--Condon} %
\newacronym{LvN}{LvN}{Liouville--von Neumann} %
\newacronym{QME}{QME}{quantum master equation} %
\newacronym{TDVP}{TDVP}{time-dependent variational principle} %
\newacronym{vMCG}{vMCG}{variational multiconfiguration Gaussians} %
\newacronym{MCTDH}{MCTDH}{multiconfiguration time-dependent Hartree} %
\newacronym{AIMS}{AIMS}{ab initio multiple spawning} %
\newacronym{NOSSE}{NOSSE}{non-stochastic open system Schr{\"{o}}dinger
  equation} %
\newacronym{SE}{SE}{Schr{\"{o}}dinger equation} %
\newglossaryentry{DM}{type=\acronymtype,name={DM},description={density
    matrix},text={DM},first={density matrix (DM)},firstplural={density
    matrices (DMs)},plural={DMs}} %
    
\title{Non-stochastic matrix Schr\"odinger equation for open systems}

\author{Lo{\"i}c Joubert-Doriol} %
\affiliation{Department of Physical and Environmental Sciences,
  University of Toronto Scarborough, Toronto, Ontario, M1C 1A4,
  Canada} %
\affiliation{Chemical Physics Theory Group, Department of Chemistry,
  University of Toronto, Toronto, Ontario M5S 3H6, Canada} %
\author{Ilya G. Ryabinkin} %
\affiliation{Department of Physical and Environmental Sciences,
  University of Toronto Scarborough, Toronto, Ontario, M1C 1A4,
  Canada} %
\affiliation{Chemical Physics Theory Group, Department of Chemistry,
  University of Toronto, Toronto, Ontario M5S 3H6, Canada} %
\author{Artur F. Izmaylov} %
\affiliation{Department of Physical and Environmental Sciences,
  University of Toronto Scarborough, Toronto, Ontario, M1C 1A4,
  Canada} %
\affiliation{Chemical Physics Theory Group, Department of Chemistry,
  University of Toronto, Toronto, Ontario M5S 3H6, Canada} %

\date{\today}

\begin{abstract}
  We propose an extension of the Schr\"odinger equation for a quantum
  system interacting with environment. This extension describes
  dynamics of a collection of auxiliary wave-functions organized as a
  matrix $\mat m$, from which the system density matrix can be
  reconstructed as $\op\rho = \mat m \mat m^\dagger$. We formulate a
  compatibility condition, which ensures that the reconstructed
  density satisfies a given quantum master equation for the system
  density.  The resulting non-stochastic evolution equation preserves
  positive-definiteness of the system density and is applicable to
  both Markovian and non-Markovian system-bath treatments. Our
  formalism also resolves a long-standing problem of energy loss in
  the \acrlong{TDVP} applied to mixed states of closed systems.
\end{abstract}


\maketitle

\glsresetall

%%%%%%%%%%%%%%%%%%%%%%%%%%%%%%
% Introduction
\section{Introduction}

Quantum evolution of a statistical mixture of
quantum states is governed by the \gls{LvN} equation $\dot{\op\rho} =
-\inumb[\op H,\op\rho]$, where $\op H$ is the Hamiltonian of a system
and $\op\rho$ is the corresponding \gls{DM}. The use of a \gls{DM} in
place of a Schr\"odinger wavefunction reflects a limited degree of
knowledge about a quantum statistical mixture. The ability of a
\gls{DM} to describe statistical phenomena becomes essential in
studies of an open quantum system interacting with its environment,
since the exact quantum state of the environment is usually impossible
to track or to control~\cite{Book/Breuer:2002}. Dynamics of an open
quantum system can be described by an appropriately modified \gls{LvN}
equation %(atomic units are used throughout)
\begin{equation}
  \label{eq:QME}
  \dt{\op\rho} = -\inumb[\op H,\op\rho] + \Nonunit[\op\rho],
\end{equation}
where a super-operator $\Nonunit$ describes non-unitary evolution of
the system due to interaction with the environment and associated
processes of relaxation, dissipation, and decoherence. We shall refer
to Eq.~\eqref{eq:QME} as the \gls{QME}. \gls{QME} can be either
postulated phenomenologically or derived from a microscopic
Hamiltonian for the open system and its environment by integrating out
environmental \gls{DOF}~\cite{Book/Breuer:2002}.

Despite its advantages, the \gls{DM}-based formalism has a few
problems. {%\color{blue}
  First, when a non-unitary part $\Nonunit$ of
  Eq.~\eqref{eq:QME} is not a generator of a completely positive map,
  the \gls{DM} may cease to be positively
  definite~\cite{Dumcke:1979/zpb/419, VanWonderen:1995/jsp/273,
    Benatti:2003/pra/042110}, which is deemed to be
  unphysical~\cite{Lindblad:1976/cmp/119, Alicki:1987} (see also the
  debates in Refs.~\onlinecite{Pechukas:1994/PRL/1060,
    Alicki:1995/PRL/3020}). In some particular cases, such as the
  Markovian dynamics, the positivity of the density matrix can be
  guaranteed by the Lindblad form~\cite{Lindblad:1976/cmp/119} of
  $\Nonunit$. However, many approximations to $\Nonunit$,
  for example, all non-Markovian ones, cannot be derived in
  that form. %
}

Second, the number of variables that are necessary to describe a pure
state $\ket{\psi}$ by a \gls{DM} as $\op\rho =\ket{\psi}\bra{\psi}$ is
a quadratic function of the Hilbert space size $\Dim_\Hilb$ for the
$\ket{\psi}$ representation. One might expect then for a mixture of
$N$ states $\Dim_\Hilb\times N$ variables at most are sufficient if an
analog of a Schr\"odinger wavefunction would exist, whereas the
\gls{DM} formalism still depends on $\sim \Dim_\Hilb^2$
variables. Thus, if $N \ll \Dim_\Hilb$, the \gls{DM} representation
seems to be excessively expensive. {%\color{blue}
  The idea to decouple
  $\Dim_\Hilb$ from $N$ has well-defined physical grounds:
  $\Dim_\Hilb$ is determined by accuracy requirements for each
  individual state, whereas the number of relevant states $N$ in an
  ensemble is dictated by a relaxation process under consideration.
  % In
  % the extreme case of relaxation to the ground state at low
  % temperature only one state eventually matters regardless of the
  % value of $\Dim_\Hilb$.
}

 Finally, a further reduction of the
Hilbert space size in quantum propagations can be achieved by using
time-evolving basis functions. The \gls{TDVP} has been successfully
used to obtain equations of motion for time-evolving basis functions
in wavefunction-based frameworks~\cite{Beck:2000/pr/1,
  Worth:2004/fd/307, BenNun:2002tx, Saita:2012di,
  Izmaylov:2013/jcp/104115}. Attempts to generalize the \gls{TDVP} to
the density formalism leads to dynamics that does not conserve the energy
of an isolated mixed-state system.~\cite{Mclachlan:1964/mp/39,
  Raab:2000/tca/358} Mathematical roots of this problem is in
conservation of the $\tr\{\op\rho^2\op H\}$ quantity by the density
\gls{TDVP} formulation instead of the system energy $E =
\tr\{\op\rho\op H\}$.

All these problems of the \gls{DM}-based formalism prompted a search
for an analog of the \gls{SE} to describe statistical ensembles of
states (mixed states). A handful of approaches of that spirit exist in
the literature. Quantum state diffusion~\cite{Diosi:1997/pla/569,
  *Strunz:1999/prl/1801, Stockburger:2001/cp/249,
  *Stockburger:2002/prl/170407} and quantum jumps~\cite{Piilo:2008di,
  Rebentrost:2009kv} methods are stochastic \gls{SE} approaches which
can describe both Markovian and non-Markovian dynamics via propagation
of wavefunction realizations. However, the propagation is stochastic,
and simulations must be converged in the number of stochastic
trajectories.  Another approach is propagation of a square-root of the
density matrix~\cite{Yahalom:2006/pa/368,*Yahalom:2007/jms/27}. This
approach was developed only for the Lindblad type of
\gls{QME}~\cite{Lindblad:1976/cmp/119}, and even within the Lindblad
scope the resulting equations of motion are not fully equivalent to
the initial \gls{QME}.  
Yet another alternative has been suggested 
through constructing an approximate, so-called surrogate Hamiltonian that 
contains system and selected environmental DOF.\cite{Koch:2002/jcp/7983,Koch:2003/jcp/1750}
Propagation of a surrogate Hamiltonian wave function allows one to obtain 
reduced system dynamics by tracing out the environmental DOF.

% Introduction of NOSSE

In what follows we introduce a formalism that generalizes the \gls{SE}
to mixed states and open systems and addresses all
  aforementioned problems. Proposed non-stochastic equations of motion
  are completely equivalent to the solution of the corresponding
  \gls{QME} provided that $\Nonunit[\op\rho]$ is a generator of a
  complete positive mapping. The equations preserve the positive
  definiteness of the density matrix by construction, and hence, the
  method may be thought of as a regularization procedure for
  non-completely positive mappings. Therefore, our formalism provides a
  very general alternative to the Lindblad form that is suitable
  for both Markonvian and non-Markovian cases. We also naturally achieve the decoupling of
  the Hilbert space size $\Dim_\Hilb$ from the number of states in an
  ensemble $N$. Finally, combining our approach with the \gls{TDVP},
  we restore the energy conservation for an \emph{isolated} mixed
  state of an arbitrary system.
%%%%%%%%%%%%%%%%%%%%%%%%%%%%%%
 
\section{Theory}

We begin by considering the system and its
environment as an isolated super-system described by a total
wavefunction $\ket{\Psi}$. Introducing a complete set of individual
states of the environment (``bath states'')
$\{\ket{\Psi_{B,s}}\}_{s=1}^\infty$, we expand the total wavefunction
as
\begin{equation}
  \label{eq:Psi_tot_exp}
  \ket{\Psi} = \sum_{s} \ket{\psi_s}\ket{\Psi_{B,s}}.
\end{equation}
Coefficients $\ket{\psi_s}=\braket{\Psi_{B,s}|\Psi}_B$ are the
functions of the system \gls{DOF} because the integration is performed
over the bath \gls{DOF} only.

The reduced density matrix $\op\rho$ of the system is obtained from
the total density matrix $\op\rho_\text{tot} =\ket{\Psi}\bra{\Psi}$ by
tracing out the bath, $\op\rho = \tr_B{\{\op\rho_\text{tot}\}}=\sum_s
\ket{\psi_s}\bra{\psi_s}$. Assuming a basis set representation for
$\ket{\psi_s}$, the reduced density can be presented in the matrix
form
\begin{equation}
  \label{eq:m}
  \op\rho = \mat m \mat m^\dagger, \quad 
  \mat m  = 
  \begin{pmatrix}
    \ket{\psi_1} & \ket{\psi_2} & \hdots
  \end{pmatrix}.
\end{equation}  
Thus, $\mat m$ is a matrix which can be thought of as a square root of
$\op\rho$. We will construct a \gls{NOSSE} for the wavefunctions $\mat
m$ in the following general form
\begin{equation}
  \label{eq:NOSSE}
  \dt{\mat m} = -\inumb\op H\mat m + \sqrtNonunit[\mat m],
\end{equation}
where $\sqrtNonunit[\mat m]$ is a functional of $\mat m$, which is
responsible for relaxation, dissipation, and decoherence due to
interaction with the environment.  The central idea of this work is
that the form of $\sqrtNonunit[\mat m]$ can be obtained exactly from
the expression for the non-unitary propagator $\Nonunit[\op \rho]$ of
the \gls{QME} in Eq.~\eqref{eq:QME}. Introducing the density
decomposition of Eq.~\eqref{eq:m} into the \gls{QME} and demanding
that Eq.~\eqref{eq:NOSSE} is satisfied, $\sqrtNonunit[\mat m]$ is
determined as a solution of the following equation:
\begin{equation}
  \label{eq:MS-Liouv}
  \sqrtNonunit[\mat m]\mat m^\dagger + \mat m\sqrtNonunit[\mat
  m]^\dagger = \Nonunit[\mat m \mat m^\dagger].
\end{equation}
Multiplication of Eq.~\eqref{eq:MS-Liouv} by $\mat m^\dagger$ and
$\mat m$ from left and right, respectively, leads to a
$\star$-Sylvester equation~\cite{Kressner:2009/nr/209}
\begin{equation}
  \label{eq:sylve}
  \mat m^\dagger\sqrtNonunit[\mat m]\left(\mat m^\dagger\mat m\right) + 
  \left(\mat m^\dagger\mat m \right) \sqrtNonunit[\mat m]^\dagger\mat m = \mat
  m^\dagger\Nonunit[\mat m\mat m^\dagger]\mat m
\end{equation}
with $\mat m^\dagger\sqrtNonunit[\mat m]$ and its Hermitian conjugate
as the new unknowns. The solution of Eq.~\eqref{eq:sylve} exists if
and only if the $\mat m^\dagger\mat m$ matrix is
invertible~\cite{Kressner:2009/nr/209}.  To recover $\sqrtNonunit[\mat
m]$ from the solution of Eq.~\eqref{eq:sylve} one needs to compute
$\mat m^{-1}$. The existence of $\mat m^{-1}$ guarantees positivity of
all eigenvalues of the density $\op\rho$.

Equations~\eqref{eq:NOSSE} and \eqref{eq:MS-Liouv} constitute a set of
coupled equations, whose solution is entirely equivalent to the
solution of \gls{QME} with a positive-defined density matrix
$\op\rho$. In many important cases one can bypass the numerical
solution of Eq.~\eqref{eq:MS-Liouv} by obtaining the form of
$\sqrtNonunit[\mat m]$ through visual inspection. For example, if
$\Nonunit[\op\rho]$ is in the Lindblad
form~\cite{Lindblad:1976/cmp/119}, $\Nonunit[\op\rho] = \sum_j 2\op
L_j \op\rho\op L_j^\dagger - \op L_j^\dagger\op L_j\op\rho -
\op\rho\op L^\dagger\op L_j$, then it is easy to check that
$\sqrtNonunit[\mat m] = \sum_j\op L_j\mat m\,(\mat m^{-1} \op L_j \mat
m)^\dagger -\op L_j^\dagger\op L_j\mat m$. Another example of the
explicit solution of Eq.~\eqref{eq:MS-Liouv} is the time
convolutionless master equation~\cite{Book/Breuer:2002} considered as
a numerical illustration below.

In fact, Eq.~\eqref{eq:MS-Liouv} admits a whole family of solutions.
If $\sqrtNonunit[\mat m]$ obeys Eq.~\eqref{eq:MS-Liouv} then
$\sqrtNonunit'[\mat m] = \sqrtNonunit[\mat m] + i\mat m \mat G$ with a
Hermitian matrix $\mat G$ also satisfies the same equation.  $\mat G$
gives rise to phase dynamics that is insignificant for any observable
properties. To show this, let us expand $\mat m$ in an orthonormal
basis set as
\begin{equation}
  \label{eq:m_exp}
  \mat m=\mat\varphi^T\mat A,  
\end{equation}
where $\mat\varphi^T = \left(\ket{\varphi_1} \ \ket{\varphi_2} \
  \hdots \right)$ is a vector of basis functions, and $\mat A$ is a
matrix of coefficients. Introducing this expansion into
Eq.~\eqref{eq:NOSSE} and considering for the simplicity a closed
system ($\Nonunit \equiv 0,\ \sqrtNonunit[\mat m] = i\mat m \mat G$),
we obtain
\begin{equation}
  \label{eq:NOSSE-trans}
  \dt{\mat\varphi}^T\mat A + \mat\varphi^T\dt{\mat A} = -i\op H
  \mat\varphi^T\mat A + i\mat\varphi^T\mat A \mat G.
\end{equation}
If we substitute $\mat A$ by $\mat A\exp(it\mat G)$, then the last
term in the right-hand side of Eq.~\eqref{eq:NOSSE-trans} will be
cancelled. The reduced density $\op\rho$ assembled from $\mat
m_G=\mat\varphi^T\mat A\exp(it\mat G)$ by Eq.~\eqref{eq:m} does not
depend on $\mat G$. Thus, the ambiguity in the solution of
Eq.~\eqref{eq:MS-Liouv} is compensated by the phase transformation of
$\mat A$ providing the $\mat G$-invariant formalism.

To solve Eq.~\eqref{eq:NOSSE} numerically, we specify $\mat m$ at
$t=0$ using the eigen-decomposition~\cite{Peschel:2012kl} of the
initial density matrix:
\begin{equation}
  \label{eq:schmidt}
  \hat\rho(0)= \sum_{j=1}^N
  \varpi_j\ket{\varphi_j(0)}\bra{\varphi_j(0)},
\end{equation}
and assigning the components of $\mat m$ to be $\ket{\psi_j} =
\sqrt{\varpi_j} \ket{\varphi_j(0)}$. In Eq.~\eqref{eq:schmidt}, $N$ is
the number of states in $\mat m$, and if each of $\ket{\varphi_j(0)}$
is described by a linear combination of $\Dim_\Hilb$ Hilbert space
vectors, then $\mat m(0)$ is a $\Dim_\Hilb\times N$ rectangular
matrix.  The inverse of a rectangular $\mat m$ is understood as the
pseudo-inverse~\cite{Moore:1920/bams/394,*Penrose:1955/mpc/406}: $\mat
m^{-1} = (\mat m^\dagger \mat m)^{-1} \mat m^\dagger$, so that the
inversion is possible if and only if Eq.~\eqref{eq:sylve} admits a
solution at time $t$ ({\it i.e.} $\mat m^\dagger\mat m$ is
invertible).  In cases when it is expected that an energy flow from
the environment will populate states $\ket{\varphi_j}$ that have small
initial weights ${\varpi_j}$ such states are added with numerically
small weights to $\mat m(0)$. When $N$ becomes larger than
$(\Dim_\Hilb+1)/2$, Eq.~\eqref{eq:MS-Liouv} appears to be
overdetermined; this is yet another manifestation of a gauge degree of
freedom associated with $\mat G$. However,
Eq.~\eqref{eq:sylve} contains the right amount of information because
the action of a rectangular $\mat m$ matrix from the left and the
right reduces the dimensionality of the matrices to the correct
values. Moreover, the freedom in the choice of $\mat G$ can be used to
reduce the number of propagated variables in $\mat m$.  If $\mat A$ at
$t=0$ in Eq.~\eqref{eq:m_exp} is brought to a lower triangular form by
applying the Cholesky decomposition to $\op\rho$ then $N$ diagonal
\emph{real} and $N(N - 1)/2$ off-diagonal \emph{complex} matrix
elements of $\mat G$ can be chosen to preserve this shape of $\mat A$
at any time. Thus, the number of coefficients to propagate is
minimized to $N\left(\Dim_\Hilb - \frac{N-1}{2}\right)$.

%%%%%%%%%%%%%%%%%%%%%%%%%%%%%%

\section{Numerical examples}

Our first example shows the \gls{NOSSE}
ability to recover the exact solution of the \gls{QME} equation for
non-adiabatic non-Markovian dynamics of an open system. We demonstrate
that our approach allows for a flexible control of the numerical
efforts needed to obtain converged results.

The model describes population transfer between two electronic states,
donor and acceptor, which are interacting with a harmonic bath.  The
system is characterized by a two-dimensional \gls{LVC} Hamiltonian
written in frequency- and mass-weighted coordinates
\begin{equation}
  \label{eq:Ham}
  \op H =  \sum_{j=1}^2\frac{\omega_j}{2}\left(\op p_j^2 + 
    \op x_j^2\right)\mat 1_2 + \left(\frac{\Delta}{2}-d\,\op
    x_1\right)\mat\sigma_z + c\op x_2\mat\sigma_x ,
\end{equation}
where $\mat 1_2$ is the $2\times2$ unit matrix, $\mat\sigma_x$ and
$\mat\sigma_z$ are the Pauli matrices. A discretized Ohmic spectral
density bath of 100 harmonic oscillators with frequencies $\Omega_k$
is coupled linearly to the $\op x_2$ coordinate with coupling
strengths $\lambda_{k}$. Numerical values of all parameters as well as
initial conditions are given in the supplemental
material~\cite{Joubert:2014un}.

The \gls{QME} is obtained using the time convolutionless approach up
to the second order in the system-bath
interaction~\cite{Book/Breuer:2002}
\begin{align}
  \label{eq:TCL-DM}
  \dt{\op\rho} & = -\inumb\left[\op H,\op\rho\right ] -
  \left(\left[\op x_2,\op x_2(t)\op\rho\right]+\left[\op\rho\op
      x_2(t)^\dagger,\op x_2\right]\right), \\
  \label{eq:correl}
  \op x_2(t) & = -\int_0^te^{-\inumb t'\op H}\op x_2e^{\inumb t'\op
    H}\sum_{j=1}^{100}\frac{\lambda_{j}^2}{2}e^{\inumb\Omega_j t'}\,
  dt'.
\end{align}
The non-unitary part $\sqrtNonunit[\mat m]$ of \gls{NOSSE} directly
follows from Eqs.~\eqref{eq:TCL-DM} and~\eqref{eq:MS-Liouv}
\begin{equation}
  \label{eq:TCL-NOSSE}
  \sqrtNonunit[\mat m] = \op x_2\mat m\left(\left[\mat m^{-1}\op
      x_2(t)\mat m\right]^\dagger-\left[\mat m^{-1}\op x_2(t)\mat
      m\right]\right).
\end{equation}

\begin{figure}
  \centering
  \includegraphics[width=0.5\textwidth]{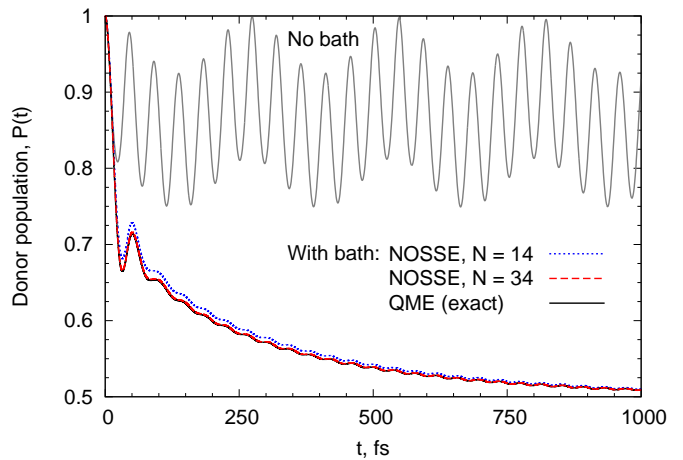}
  \caption{Dynamics of the donor state population $P(t) = \tr\{
    \op\rho(t)[\mat 1+\mat\sigma_z]\}/2$ starting from the Boltzmann
    distribution of the donor state at $T = 1000$\,K. The bath is at
    the Boltzmann distribution with $T = 0$\,K.}
  \label{fig:pop-1}
\end{figure}

To simulate converged population dynamics (Fig.~\ref{fig:pop-1}) the
\gls{QME} approach employed $240$ basis functions, which correspond to
$28920$ independent matrix elements in $\op \rho$. As few as $N=14$
states in $\mat m$ provided a very good agreement between \gls{NOSSE}
and \gls{QME} calculations, and with $N = 34$ the results are
converged. This leads to only $3269$ and $7599$ unique matrix elements
in \gls{NOSSE} to propagate for $N = 14$ and $N = 34$,
respectively. It is important to note that 
these reductions of the number of matrix elements 
will exponentially grow with the number of system DOF for larger systems.
To choose $\mat m$ states we use not only the population
criterion in the initial density eigen-decomposition but also the
system and environment energy scales corresponding to the model
dynamics~\cite{Joubert:2014un}.

Although there is a clear reduction of the number of propagated variables in the NOSSE formalism, it is important 
to confirm that this reduction leads to higher computational efficiency in terms of CPU time because 
there can be other factors (e.g., integration time step) that might reduce efficiency of NOSSE as compared to that of QME. 
Equation (\ref{eq:TCL-DM}) is solved by the Runge--Kutta--Fehlberg (RKF45) algorithm as implemented in the 
ode45 Matlab function~\cite{Matlab:2012b} on a twelve-core node Intel(R) Xeon(R) CPU E5-2630 @ 2.30GHz with $64$ GB of memory
and dynamical allocation of number of cores. 
CPU times for simulations in Fig.~\ref{fig:pop-1} are given in Table~\ref{tab:2} (non-Markovian) 
and show insignificant gain from using NOSSE.
However, a closer examination reveals that evaluation of the non-Markovian correlator 
in Eq.~(\ref{eq:correl}) at each time step takes almost $98\%$ of all CPU time, and it is the same 
for both QME and NOSSE formalisms.
To perform a more adequate comparison we simulate the same propagation 
using the Markovian approximation [$t\rightarrow\infty$ in Eq.~(\ref{eq:correl})]
to avoid calculating the correlator at each time step. 
In this case, \gls{NOSSE} is faster than \gls{QME} by a factor of $2.2$ (Table~\ref{tab:2}).
\begin{table}[!b]
  \centering
  \caption{CPU time (min) for the first $100$ femtoseconds of propagation. \protect\gls{NOSSE} values are for $N=34$ states.}
  \begin{ruledtabular}
    \begin{tabular}{@{}l@{}c@{}c@{}}
      Correlator in Eq.~(\ref{eq:correl}) &   \gls{QME}    &   \gls{NOSSE}  \\ \hline  
      Non-Markovian &  802  &  797  \\
      Markovian, $t\rightarrow\infty$     &  29  &  13
    \end{tabular}
  \end{ruledtabular}
  \label{tab:2}
\end{table}

%%%%%%%%%%%%%%%%%%%%%%%%%%%%%%%%%%%%%%%%%%%%%%%%%%%%%%%%%%%%%%%%%%%%%%
% Energy conserved vMCG for density matrix
Our second example shows how the \gls{NOSSE} formalism combined with
the \gls{TDVP} leads to a set of equations of motion that conserve the
energy of an \emph{isolated} system.  The \gls{TDVP} allows one to
derive approximate equations of motion when basis functions \emph{and}
their coefficients are time-dependent~\cite{Mclachlan:1964/mp/39}. For
the \gls{DM} formalism the \gls{TDVP} amounts to minimization of the
Hilbert-Schmidt norm of an operator $\big\lVert\dt{\op\rho}+\inumb[\op
H,\op\rho]\big\rVert$. In the \gls{NOSSE} formalism we minimize
${\big\lVert\dt{\mat m} +\inumb\op H\mat m\big\rVert}$.

\begin{figure}
  \centering
  \includegraphics[width=0.5\textwidth]{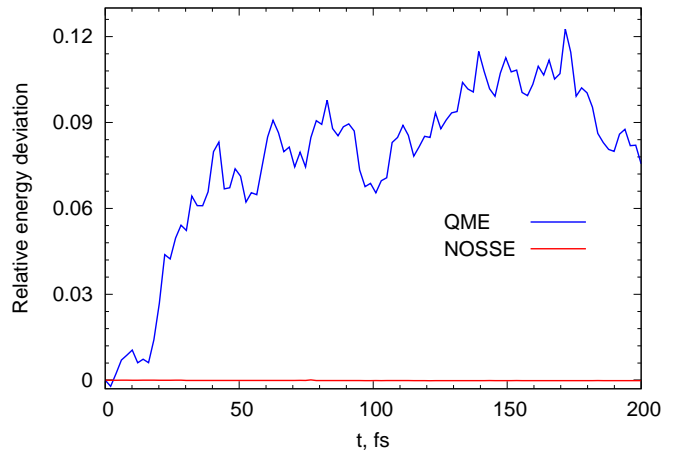}
  \caption{Relative energy deviation ${\big(\tr{\{\op
        H\op\rho(t)\}}-E_0\big)/{\omega_1}}$ for dynamics based on the
    \protect\gls{QME} [Eqs.~\eqref{eq:vMCG-DM-C}
    and~\eqref{eq:vMCG-DM-Y}] and \protect\gls{NOSSE}
    [Eqs.~\eqref{eq:vMCG-MS-C} and~\eqref{eq:vMCG-MS-Y}] formalisms. A
    basis set of $41$ Gaussians is used in both cases.}
  \label{fig:energy-2}
\end{figure}

Consider an isolated system described by the Hamiltonian in
Eq.~\eqref{eq:Ham}. To apply the TDVP, we employ a linear combination
of Gaussian products
\begin{align}
  \label{eq:anzatz}
  \op\rho & = \sum_{j,k=1}^{N_g}
  \ket{g_j}B_{jk}\bra{g_k}, \\
  \braket{x_1 x_2|g_j} & = \exp{\left[\sum_{\alpha =
        1}^{2}\left(-\frac{K_{j,\alpha} x_\alpha^2}{2} +
        \xi_{j,\alpha}x_\alpha\right)\right]}.
\end{align}
The widths of Gaussian functions are fixed $K_{j,\alpha} = 1$, whereas
$\mat B = \{B_{jk}\}$ and $\mat\xi = \{\xi_{j,\alpha}\}$ are the
time-dependent quantities to be optimized via the
\gls{TDVP}~\cite{Izmaylov:2013/jcp/104115}.  Minimization of
$\big\lVert\dt{\op\rho}+\inumb[\op H,\op\rho]\big\rVert$ results in
the following equations of motion~\cite{Burghardt:1999/jcp/2927}
\begin{eqnarray}
  \label{eq:vMCG-DM-B}
  \dt{\mat B} & = & -\mat S^{-1}\left(\inumb\mat H+\mat\tau\right)\mat
  B + \mat B\left(\inumb\mat H-\mat\tau^\dagger\right)\mat S^{-1}, \\
  \label{eq:vMCG-DM-lambda} 
  \dt{\mat\xi} & = & \mat C^{-1}\mat Y,
\end{eqnarray}
where $S_{kl} = \braket{g_k|g_l}$, $\tau_{kl} =
\braket{g_k|\frac{\partial g_l}{\partial t}}$, $H_{kl} =
\braket{g_k|\op H|g_l}$,
\begin{eqnarray}
  \label{eq:vMCG-DM-C}
  C^{\alpha \beta}_{kl} & = & \braOket{\frac{\partial g_k}{\partial
      \xi_{k,\alpha}} } { \left[\op 1 - \op P_{N_g} \right] } 
  { \frac{\partial g_l}{\partial\xi_{l,\beta}} } [\mat B\mat S\mat
  B]_{lk}, \\
  \label{eq:vMCG-DM-Y}
  {[\mat Y]}^\alpha_k & = & -\inumb\sum_l\braOket{\frac{\partial
      g_k}{\partial \xi_{k,\alpha}} }{\left[\op 1 - \op
      P_{N_g}\right]\op H\op\rho}{g_l}B_{lk}, 
\end{eqnarray}
and $\op P_{N_g} = \sum_{mn}\ket{g_m}[\mat S^{-1}]_{mn}\bra{g_n}$.  On
the other hand, applying the \gls{TDVP} to \gls{NOSSE} recovers
Eqs.~\eqref{eq:vMCG-DM-B} and \eqref{eq:vMCG-DM-lambda} with new $\mat
C$ and $\mat Y$:
\begin{eqnarray}
  \label{eq:vMCG-MS-C}
  C^{\alpha\beta}_{kl} & = & \braOket{\frac{\partial g_k}{\partial
      \xi_{k,\alpha}} } { \left[\op 1 - \op P_{N_g} \right] } 
  { \frac{\partial g_l}{\partial\xi_{l,\beta}} } B_{lk}, \\
  {[\mat Y]}^\alpha_k & = & -\inumb\sum_l\braOket{\frac{\partial
      g_k}{\partial \xi_{k,\alpha}} }{\left[\op 1 - \op
      P_{N_g}\right]\op H}{g_l}B_{lk}.
  \label{eq:vMCG-MS-Y}
\end{eqnarray}
To expose the problem of energy conservation we used a different set
of parameters and initial conditions than in our first
example~\cite{Joubert:2014un}. The total energy of the system as a
function of time is plotted in Fig.~\ref{fig:energy-2}. It is clear,
that TDVP equations based on \gls{NOSSE} definitions
[Eqs.~\eqref{eq:vMCG-MS-C} and~\eqref{eq:vMCG-MS-Y}] conserve the
total energy of an isolated system, whereas the \gls{DM} counterparts
[Eqs.~\eqref{eq:vMCG-DM-C} and~\eqref{eq:vMCG-DM-Y}] do not.

%%%%%%%%%%%%%%%%%%%%%%%%%%%%%%

\section{Conclusions}

We introduced a non-stochastic analog of the
Schr\"odinger equation for open systems, NOSSE, that involves a set of wave
functions.  The equations of motion for this set are made to reproduce
the \gls{QME} evolution for the system density.  Our formalism
guarantees the positivity of the \gls{DM}
for any propagation time (as long as the solution exists) and it is applicable
to Markovian and non-Markovian treatments of system-bath interaction.
\gls{NOSSE} is equivalent to \gls{QME} only if the latter generates a positive mapping. 
In cases when \gls{QME} does not guarantee the \gls{DM} positivity,
\gls{NOSSE} enforces the positivity through a regularization procedure.
Although the restoration of the \gls{DM} positivity seems very attractive, 
one cannot guarantee that this feature will always improve the solution in a 
sense of the norm difference between approximate and exact solutions. 
Nevertheless, \gls{NOSSE} is a general formalism that 
can operate with dissipators that do not suffer from loss of the \gls{DM} positivity.
From a numerical standpoint, our formalism is more
computationally efficient since the number of dynamical variables is
no longer quadratic with respect to size of the Hilbert space. The new
approach is compatible with the \gls{TDVP}, since it leads
to total energy conservation for closed systems, and can be applied in
investigating photochemical reactions induced by incoherent light.

%%%%%%%%%%%%%%%%%%%%%%%%%%%%%%
% Acknowledgements
\section{Acknowledgements}

LJD thanks the European Union Seventh Framework Programme
(FP7/2007-2013) for financial support under grant agreement
PIOF-GA-2012-332233. AFI acknowledges funding from the Natural
Sciences and Engineering Research Council of Canada (NSERC) through
the Discovery Grants Program.

%\bibliography{NOSSE}
%

\end{document}